\documentclass[aps,prl,twocolumn,superscriptaddress]{revtex4-1}

\usepackage{hyperref}
\usepackage{graphicx}

\newcommand{\csaui}{Cs$_2$Au$_2$I$_6$ }
\begin{document}


\title{Short small-polaron lifetime in the mixed-valence perovskite Cs$_2$Au$_2$I$_6$ from high-pressure pump-probe experiments}

\author{M. Trigo}
\email[]{mtrigo@slac.stanford.edu}

\author{J. Chen}

\author{M. P. Jiang}
\affiliation{Stanford PULSE Institute, SLAC National Accelerator Laboratory, Menlo Park, CA 94025, USA}
\author{W. L. Mao}
\affiliation{Department of Geological \& Environmental Sciences, Stanford University, Palo Alto, CA, USA}
\author{S. C. Riggs}
\affiliation{Department of Applied Physics and Geballe Laboratory for Advanced Materials, Stanford University, Stanford, CA, USA}
\author{M. C. Shapiro}
\affiliation{Department of Applied Physics and Geballe Laboratory for Advanced Materials, Stanford University, Stanford, CA, USA}
\author{I. R. Fisher}
\affiliation{Department of Applied Physics and Geballe Laboratory for Advanced Materials, Stanford University, Stanford, CA, USA}
\author{D. A. Reis}
\affiliation{Stanford PULSE Institute, SLAC National Accelerator Laboratory, Menlo Park, CA 94025, USA}
\affiliation{Department of Photon Science and Applied Physics, Stanford University, Stanford, CA 94305, USA}

\date{\today}

\begin{abstract}
We study the ultrafast phonon response of mixed-valence perovskite \csaui using pump-probe spectroscopy under high-pressure in a diamond anvil cell. We observed a remarkable softening and broadening of the Au - I stretching phonon mode with both applied pressure and photoexcitation.
Using a double-pump scheme we measured a lifetime of the charge transfer excitation into single valence Au$^{2+}$ of less than $4$~ps, which is an indication of the local character of the Au$^{2+}$ excitation.
Furthermore, the strong similarity between the pressure and fluence dependence of the phonon softening shows that the inter-valence charge transfer plays an important role in the structural transition.
\end{abstract}


\maketitle

Mixed valence (MV) materials based on $4f$ elements have historically attracted considerable attention\cite{Robin1968,Varma1976}. 
In materials such as the samarium chalcogenides as well as in many cerium intermetallic compounds, the valence fluctuates rapidly between $f^n$ and $f^{n-1}$ but the average valence is nevertheless homogeneous\cite{Coey1976,Annese2006}.
Driven by hybridization with itinerant carriers, this behavior is intimately related to the Kondo effect\cite{Hewson}. A distinctly different type of mixed valence occurs for some $d$-block elements for which Jahn-Teller effects stabilize static (inhomogeneous) $d^{n-1}$ and $d^{n+1}$ configurations\cite{Robin1968}. Much less is known about valence fluctuations in this class of material, for which coupling to the crystal lattice clearly plays an important role. 
Here we study Cs$_2$Au$_2$I$_6$, which is a canonical example of a $5d$ mixed valence system exhibiting charge disproportionation\cite{kojimaBCSJ2000}. Of particular interest, this material undergoes a first order coupled structural and valence transition at a moderately low pressure of just 5.5 GPa\cite{Kojima_J.Am.Chem.Soc1994,Matsushita1997}. Using ultrafast pump-probe spectroscopy we make a quantitative comparison of the effects of photoinduced valence transitions with those of hydrostatic pressure, as revealed by softening of the phonon spectra. The clear similarity between these two cases confirms that the structural transition is associated with the inter-valence charge-transfer (IVCT) between the distinct Au sites. The photoinduced state has a remarkably short lifetime, indicative of local (small) polaronic excitations rather than extended states.

\begin{figure}
\includegraphics[scale=0.32]{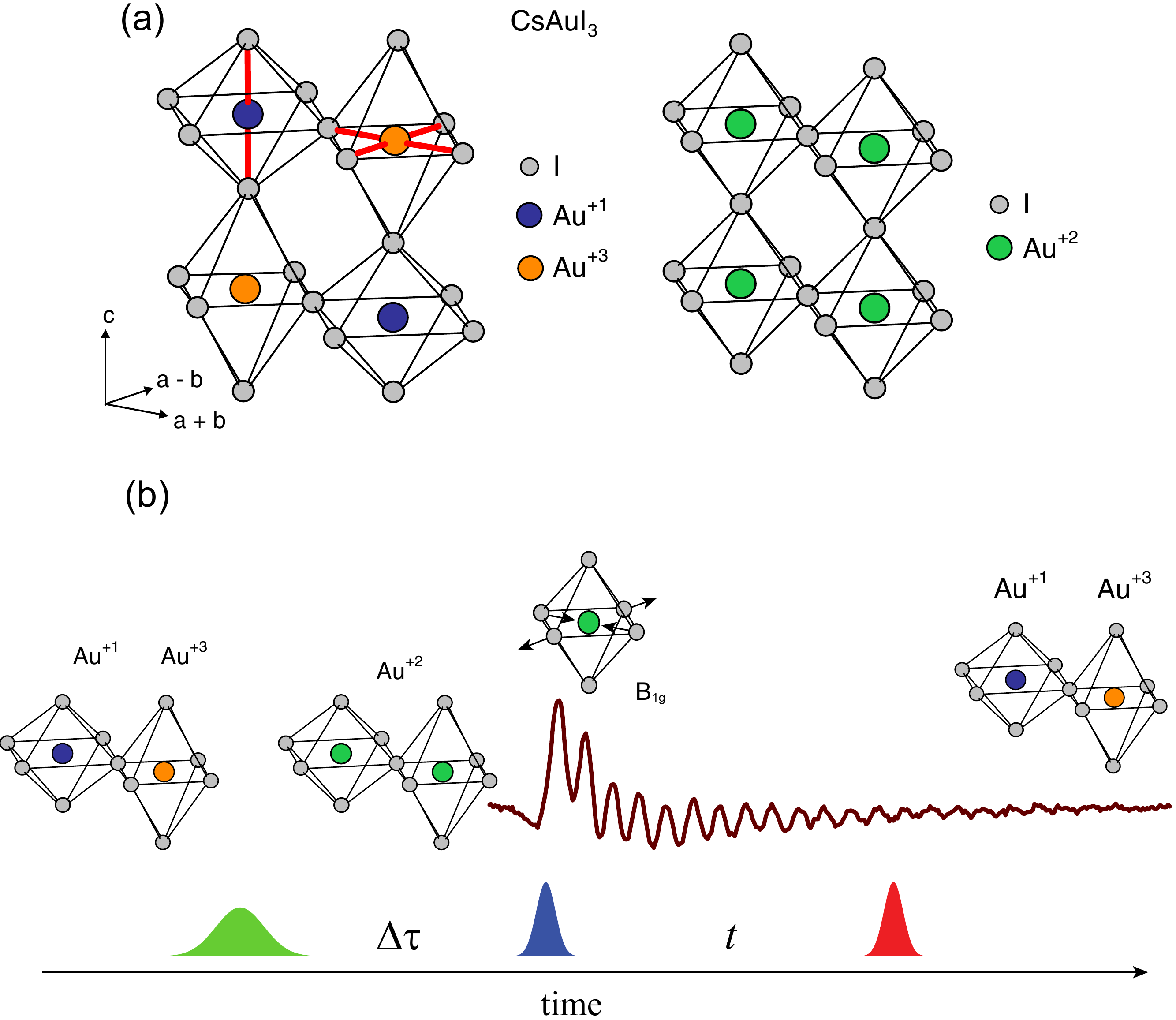}%
\caption{(a) The structure of \csaui in the mixed valence (left) and single valence (right) states with the accompanying band Jahn-Teller lattice distortion  [for clarity the Cs atoms are not displayed]. (b) Schematic diagram of the three pulse scheme for high-pressure pump-probe spectroscopy discussed in the text: a stretch pulse (green) induces the IVCT Au$^{1+} \to$  Au$^{3+}$, a delayed short pulse (blue) excites coherent modes and a third pulse (red) probes the reflectivity changes due to the phonon.
\label{Fig1}}
\end{figure}

The Cs$_2$Au$_2$X$_6$ family, with $X = $ Cl, Br or I, is a mixed-valence perovskite that shows a pressure induced phase transition at $11$, $9$ and $5.5$~GPa as the size of the halogen atom increases from Cl, to Br to I, respectively\cite{kojimaBCSJ2000,Liu1999,Matsushita2007}. In the low pressure phase, the gold ion appears in a MV state with alternating Au$^{1+}$ and Au$^{3+}$ sites and the surrounding halogen ions form a distorted octahedron as depicted in Fig.~\ref{Fig1} (a). 
At the critical pressure the system transforms from MV to single-valent (SV) and all the Au ions acquire a valence $2+$\cite{Hafner1994}. At this point the surrounding halogen ions shift to the symmetric position between neighboring Au$^{2+}$ [Fig.~\ref{Fig1} (a)] and the associated lattice distortion disappears\cite{kojimaBCSJ2000}. As in the insulator BaBiO$_3$, the strong coupling between the charge disproportionation of the metal ion and the lattice deformation stabilizes the commensurate charge density wave (CDW).

Pump-probe experiments were performed using pulses from a Ti:sapphire regenerative amplifier laser (Coherent Rega) with nominal pulse duration of $50$~fs and a central wavelength of $800$~nm.
After the amplifier, $50~\%$ of the beam was split to seed an optical parametric amplifier (OPA) which provides pulses with $\sim 0.12~\mu$J of energy at a central wavelength of $550$~nm. The other $50\%$ of the beam was split into probe ($800$~nm) and pump ($400$~nm) which was frequency-doubled on a nonlinear beta barium borate (BBO) crystal [see Fig. \ref{Fig1} (b) for a diagram of the three pulse sequence used later in the text].
Depending on the configuration of the experiment (see discussion below), a combination of pulses with wavelengths $800$, $550$ and $400$~nm were overlapped to a spot $\sim 10~\mu$m in diameter inside a diamond anvil cell (DAC) with a microscope objective [see Fig.~\ref{Fig1} (b)]. The back-scattered light was refocused and spatially filtered with a $100~\mu$m pinhole, which was then imaged onto a Si photodiode. A color filter in front of the detector rejected scattered $400$~nm pump light.
The $400$~nm pump was chopped at $1.6$~kHz and a lock-in scheme was used to detect the differential reflectivity at $800$~nm. The probe fluence was kept below $0.3$~mJ/cm$^2$.
Single crystals of ${\rm Cs_2Au_2I_6}$ were prepared by a self flux technique, as described elsewhere\cite{Riggs2011}. Samples were crushed in the DAC in order to ensure that backscattered light reached the detector. The gasket hole was $200~\mu$m in diameter and silicone oil was used as the pressure medium.  The pressure was calibrated by monitoring the fluorescence of a small ruby crystal inside the DAC. 

\begin{figure}
\includegraphics[scale=0.4]{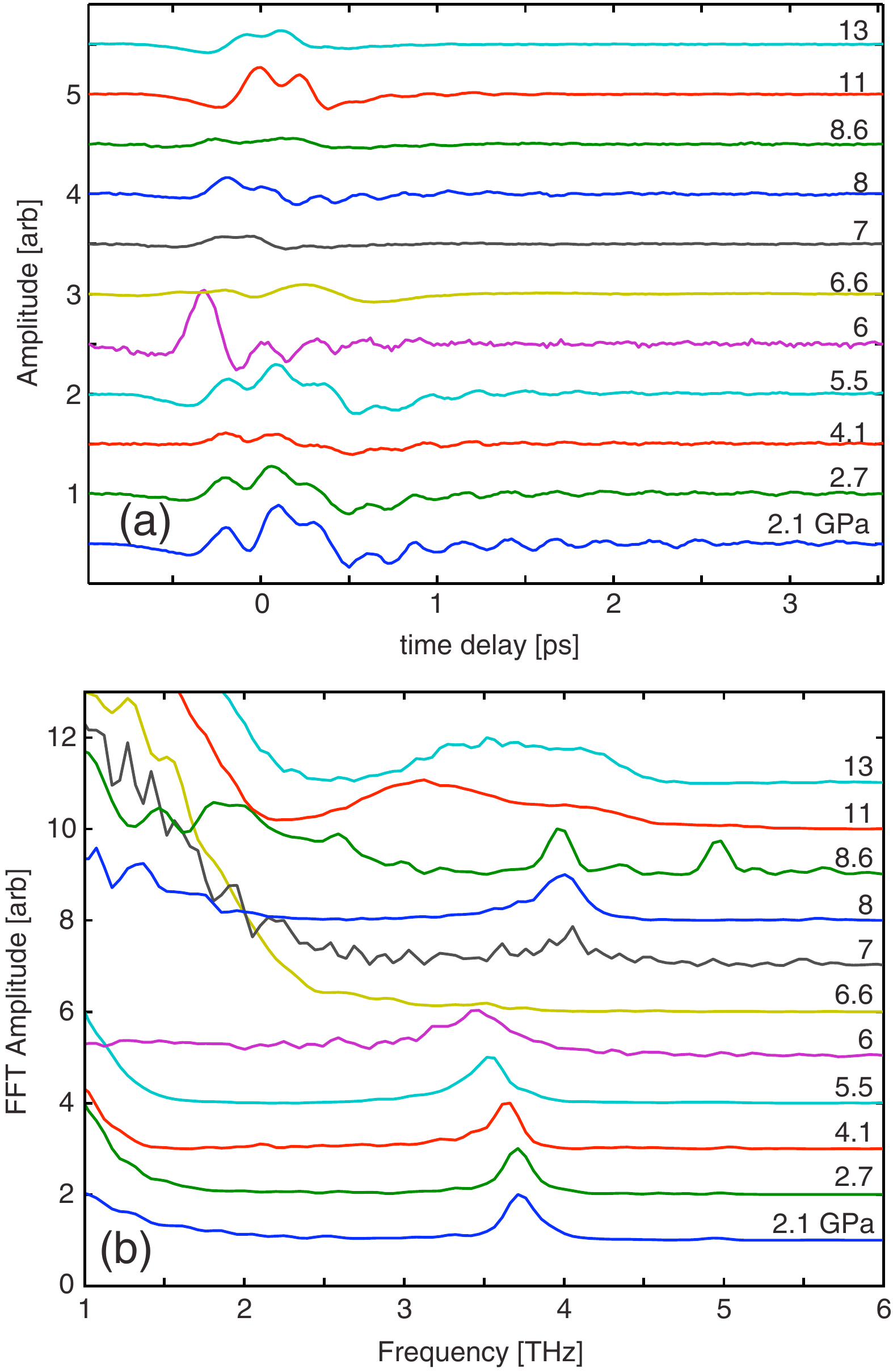}%
\caption{(a) Oscillatory component of the differential reflectivity for different hydrostatic pressures (listed to the right of each trace). (b) Fourier transforms of the traces in  (a). The phonon mode near $3.6$~THz disappears at $\sim 6.6$~GPa and oscillations with a frequency of $4$~THz appear above the transition around $8$~GPa.
\label{Fig2}}
\end{figure}

Figure \ref{Fig2} shows pump- ($400$~nm) probe ($800$~nm) traces for varying pressure (a) and their Fourier transforms (b). Curves in (a) were filtered by subtracting a moving average to remove the slowly varying background. 
All the traces were taken with the same pump fluence $\sim 0.3$~mJ/cm$^2$, which is low enough to avoid softening the phonon mode significantly by photoexcitation. At the lowest pressure, the phonon oscillations appear close to $3.7$~THz, as seen in the Fourier transform [Fig.~\ref{Fig2} (b) lowest trace]. 
This oscillation is assigned to the $B_{1g}$ stretching mode of the distorted $I_6$ octahedron surrounding the Au atoms. The displacement associated with this mode is schematically drawn in Fig.~\ref{Fig1} (b).
As pressure approaches the phase transition near $6$~GPa the frequency softens and the peak broadens and eventually disappears at $\sim 6.6$~GPa [Fig.~\ref{Fig2} (b)]. 
Similar softening behavior due to the freezing of the Jahn-Teller-like modes has been observed at room temperature in the Raman spectra of ${\rm Cs}_2{\rm Au}_2{\rm Br}_6$ and ${\rm Cs}_2{\rm Au}_2{\rm Cl}_6$ at $9$ and $11$~GPa, respectively\cite{Liu1999}. The Raman intensity of the $B_{1g}$ mode vanishes above the transition due to the doubling of the Brillouin zone, which unfolds the zone-center $B_{1g}$ mode into the edge of the new Brillouin zone, making this mode inaccessible to Raman scattering due to its large wavevector\cite{Loudon1964}.
Above the transition ($P > 7$~GPa), different oscillations appears at $> 4$~THz, which must originate from different zone-center modes of the high-pressure phase.
Eventually these oscillations also disappear above $\sim 13$~GPa where the material may become amorphous\cite{Kusmartseva2010}.

Many of the physical properties of \csaui can be understood by thinking of the system as being composed of alternating linear [Au$^{1+}$I$_2]^-$ and square planar [Au$^{3+}$I$_4]^-$ molecules\cite{Matsushita1997}, as represented by the red nearest-neighbor bonds in Fig.~\ref{Fig1} (a). The observed oscillations can be ascribed to the coherent motion of the $B_{1g}$  mode of the square [Au$^{3+}$I$_4]^-$ molecule\cite{kojimaBCSJ2000}. 
At ambient pressure we observed a weak peak at $4.6$~THz that we assign to the higher-frequency $A_{1g}$ mode of the linear [Au$^{1+}$I$_2]^-$ molecule. However, this mode vanishes at lower pressure than the $B_{1g}$ mode\cite{Liu1999} and was not visible at $P > 2$~GPa.
At $3.5$~GPa, ${\rm Cs}_2{\rm Au}_2{\rm I}_6$ is in the mixed-valence state with alternating Au$^{1+}$ and Au$^{3+}$ cations [Fig.~\ref{Fig1} (a)]. As the pressure increases above $\sim 5.5$~GPa, increased overlap between the molecular orbitals of the gold iodide complexes is thought to drive the system into the single Au$^{2+}$ valence. The suppression of the Au$^{1+} - {\rm Au}^{3+}$ charge order relaxes the octahedral distortion\cite{kojimaBCSJ2000, Matsushita2007}, and results in the softening and broadening of the phonon mode as seen in Fig.~\ref{Fig2} (b). 
Significantly, the IVCT between the $5 d$ orbitals can also be achieved by photo-excitation in the visible region of the spectrum\cite{kojimaBCSJ2000}.
As we show next, photoexcitation has a similar effect on the phonon response as pressure, and argue that there is a direct analogy between the effect of pressure and photo-excitation on the system since both induce the IVCT. 

\begin{figure}
\includegraphics[scale=0.3]{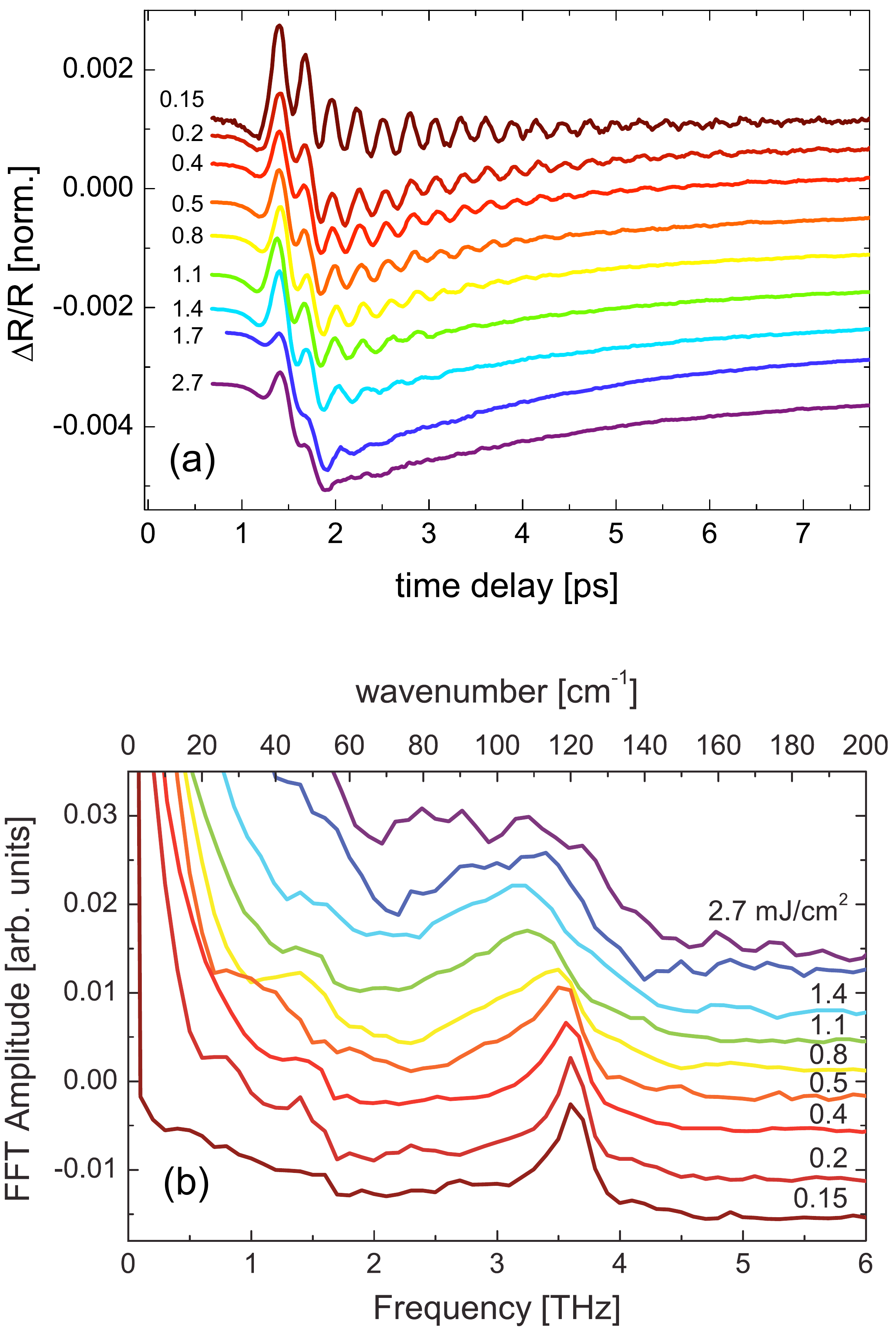}%
\caption{(a) time-dependence of the differential reflectivity for different $400$~nm pump fluences at $3.5$~GPa. The traces are normalized by their maximum amplitude for clarity. Incident fluences in mJ/cm$^2$ are shown next to each trace. (b) representative Fourier transforms of time-traces for different pump fluences (normalized).
\label{Fig3}}
\end{figure}

Figure \ref{Fig3} (a) shows single- pump- (400 nm) probe (800 nm) differential reflectivity traces at $\sim 3.5$~GPa for increasing pump fluences. 
The traces were normalized to their maximum value for displaying purposes (however their amplitude is linear with fluence). Figure~\ref{Fig3} (a) shows coherent oscillations similar to those in Fig.~\ref{Fig2}, which correspond to the same $B_{1g}$ stretching mode of the $I_6$ octahedron sketched in Fig.~\ref{Fig1} (a). The oscillation in the lowest fluence trace in Fig. \ref{Fig3} (a) (top trace) shows a slight frequency chirp that starts at $3.61$~THz ($277$~fs period) measured at the first oscillation cycle that becomes $3.74$~THz ($267$~fs period) at the 6$^{\rm th}$ cycle. The softening induced by the absorption of the pump pulse is more evident in the Fourier transforms of these traces as shown in Fig. \ref{Fig3} (b), where the peak at $\sim 3.6$~THz softens and broadens as the pump fluence is increased. At higher excitations the phonon oscillation was less visible due to the quickly decaying phonon amplitude  and the large background. Note that the maximum fluence $2.7$~mJ/cm$^2$ corresponds to $\sim 5\times 10^{15}$~photons/cm$^2$, well below the nucleation threshold for complete transformation to single valence reported for ${\rm Cs}_2{\rm Au}_2{\rm Br}_6$\cite{LiuPRB2000}.
The picture that emerges from the pressure and fluence dependence in Fig.~\ref{Fig2} and \ref{Fig3} is the following:  As pressure increases, a gradual overlap between the $5d_{x^2-y^2}$ orbitals of Au$^{1+}$ and Au$^{3+}$ due to the crystal-field modification induces the transition to the single-valence Au$^{2+}$  at $5.5$~GPa\cite{kojimaBCSJ2000}. 
In a completely analogous way, optical excitation induces the IVCT Au$^{1+} \to$  Au$^{3+}$ suppressing the charge modulation, which results in a softening and deactivation of the phonon. 
In this discussion, the short pump pulse plays two roles: it excites the coherent $B_{1g}$ motion impulsively\cite{Merlin1997}, which leads to the modulation of the optical reflectivity in Fig.~\ref{Fig3} (a), and at the same time produces the IVCT Au$^{1+} \to$  Au$^{3+}$, which induces the observed softening of the phonon mode. 

In order to gain a better understanding of the photoexcited state it is useful to separate the optical IVCT from the impulsive excitation of the phonon. To do this, we implemented a three-pulse pump-probe scheme depicted schematically in Fig.~\ref{Fig1} (b): the first pulse induces the electronic charge transfer (green pulse, 550 nm). This photon energy is close to the $5 d_{xy} \to 5 d_{x^2-y^2}$ transition between Au$^{1+}$ and Au$^{3+}$\cite{LiuPRB1999}, which ``prepares'' the system into a significant fraction of Au$^{2+}$ sites. The pulse duration was stretched to $600$~fs to avoid generating coherent phonons by impulsive excitation\cite{Merlin1997} with the ``preparation'' pulse.
A delayed short pulse (blue, 400 nm) impulsively excites the coherent motion of the phonon of the IVCT state, which is softened by the photoinduced homogeneous charge distribution. 
Finally, a third weak pulse (red, 800 nm) with a different optical delay probes the differential optical reflectivity. We define $\Delta \tau$ as the delay between the IVCT pulse (550~nm) and the phonon pump (400~nm), while ``probe delay'' is the usual delay between the coherent phonon pump (400~nm) and the probe pulse (800~nm).
In this way, with independent control of the delays we measure the impulsive phonon response of the intermediate state and use it as a probe of the dynamics of the IVCT excitation.

\begin{figure}
\includegraphics[scale=0.4]{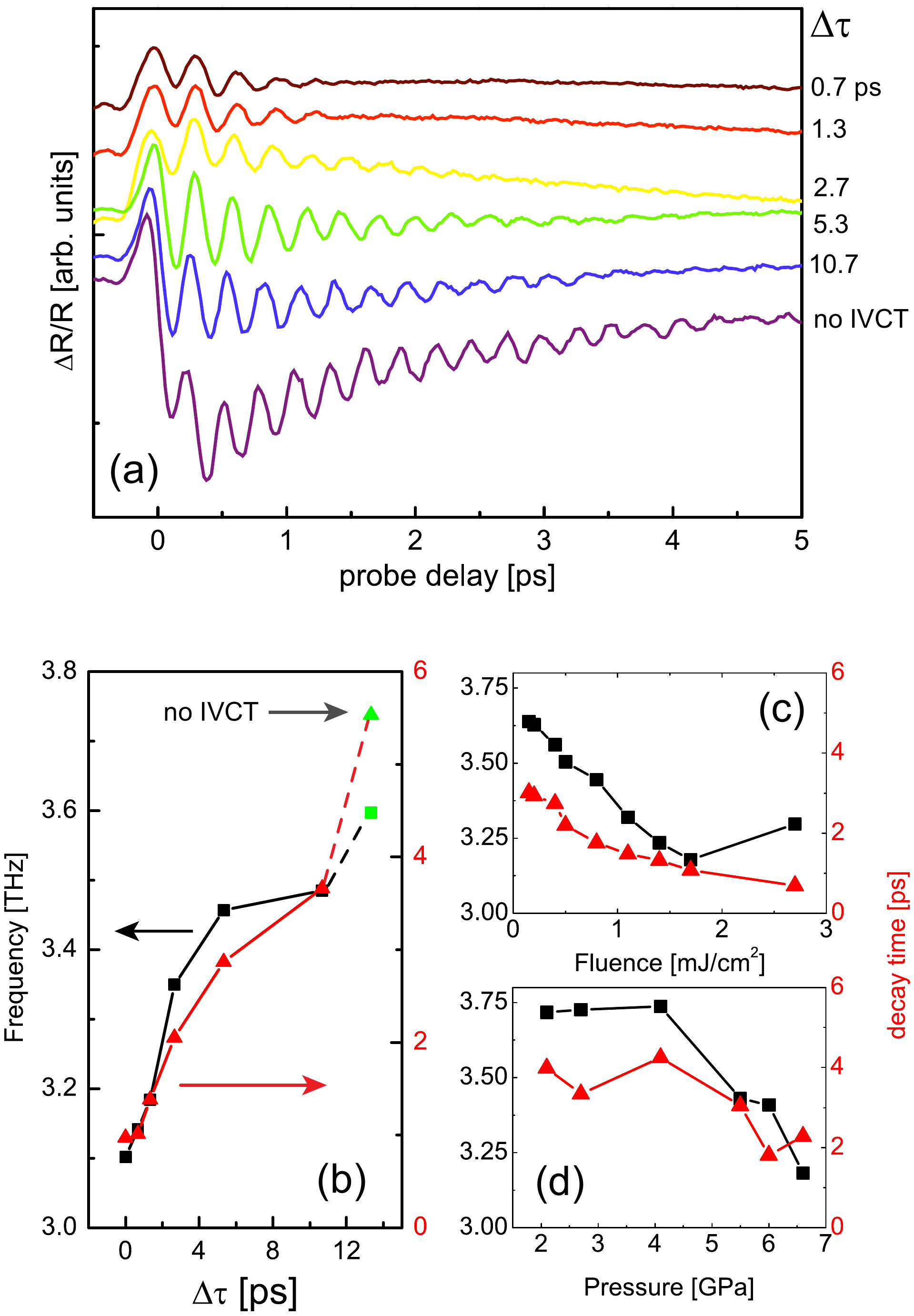}%
\caption{(a) Coherent phonon response at 3.5 GPa for different delays $\Delta \tau$ after charge-transfer excitation with $550$~nm pulses. (b) Frequency and decay constant of the coherent oscillations in (a) as a function of charge transfer pump delay $\Delta \tau$. (c) Fluence and (d) pressure dependence of the frequency and decay constant of the oscillations from Figs~\ref{Fig2} and \ref{Fig3}. 
\label{Fig4}}
\end{figure}

Figure \ref{Fig4} (a) shows time-resolved differential reflectivity vs. probe delay for different $\Delta \tau$. The traces resemble the single-pump data in Fig. \ref{Fig3} (a) with coherent oscillations due to the $B_{1g}$ phonon mode around $3.7$~THz. However, here each curve represents the phonon response to an impulsive force at time $\Delta \tau$ after the IVCT ``preparation'' pulse (i.e., $\Delta \tau$ after creation of Au$^{2+}$ sites).
As with the fluence data in Fig.~\ref{Fig3}, the oscillations are softened and strongly damped, particularly for small $\Delta \tau$, i.e. immediately after the IVCT excitation [top trace in Fig.~\ref{Fig4} (a)]. As $\Delta \tau$ increases [Fig.~\ref{Fig4} (a) top to bottom] the IVCT relaxes back to equilibrium and the phonon response recovers towards the single-pump trace labeled ``no IVCT''.
The $400$~nm fluence was kept low to avoid additional softening of the mode.
Figure \ref{Fig4} (b) shows the frequency and decay constant of the oscillations in (a) vs. $\Delta \tau$ extracted by fitting the oscillatory component of the time trace with a decaying cosine function. 
For comparison, figures~\ref{Fig4} (c) and (d) show the fluence and pressure dependence of these parameters from the data in figures \ref{Fig2} and \ref{Fig3} [on the same scale as in (b)].
Figure \ref{Fig4} (b) brings out more clearly the softening due to the IVCT: the frequency (black squares) drops to almost $85~\%$ of the nominal value near $\Delta \tau = 0$ while at $\Delta \tau \sim 4$~ps it has already recovered to $> 97~\%$ of  the value for the unperturbed phonon (green symbols). 
This represents a measurement of the lifetime of the IVCT excitation. Such a fast recovery time for the charge is unusual for typical extended states in solids which can range up to $\mu$s in some semiconductors, and is likely due to the local nature of the charge excitation associated with the induced Au$^{2+}$ small-polaron sites\cite{Kusmartseva2010}.
The decay constant also varies drastically with $\Delta \tau$, fluence, and pressure [red triangles in Fig.~\ref{Fig4} (b), (c) and (d)]. 
This behavior can be explained by an inhomogeneous broadening caused by the presence of single and mixed valence clusters of varying sizes\cite{LiuJPSJ2000}.
Indeed, a Gaussian distribution of decaying oscillators with variance $\sigma = 0.5$~THz gives $\tau_{1/e}$=2.2 ps assuming an intrinsic oscillator linewidth of $\Gamma = 1/6$~[ps$^{-1}$] (estimated from the decay of the unperturbed ``no IVCT'' trace). These values are consistent with the range of decay times observed in Figs.~\ref{Fig4} (b), (c) and (d), indicating that the decay in the signal is mostly due to inhomogeneous broadening. Note however that we cannot rule out a contribution to the broadening due to inhomogeneous pump illumination. 
Finally, the similarity between the pressure and fluence dependence in (c) and (d) is consistent with a picture in which the phonon softening and thus the structural transition is intimately connected with the charge transfer excitation. 

In conclusion, we studied the ultrafast phonon response of \csaui in-situ using pump-probe under high-pressure inside a DAC. We observed a remarkable softening and broadening of the $B_{1g}$ Au - I stretching phonon mode with both applied pressure and photoexcitation. This mode involves motion of the distorted $I_6$ octahedron surrounding the Au ion, and thus is strongly coupled with the Au valence electrons. We observe a strong similarity between the pressure and fluence dependence of the phonon response, which is consistent with a picture in which the IVCT and the structural transition are intimately connected.
Furthermore, using a three-pulse pump-probe sequence we measured a lifetime of the IVCT excitation of $< 4$~ps. We argue that such a fast recovery of the valence is an indication of the local (small-polaron) character of the  Au$^{2+}$ excitation, which also provides compelling evidence for the system being non-metallic.
These results also demonstrate the utility of pump-probe spectroscopy to probe the dynamics of excitations in correlated matter under high-pressure.

MT, JC, MPJ, WLM and DR acknowledge support from the US Department of Energy office of Basic Energy Science through the PULSE and SIMES institutes. 
SCR, MS and IRF are supported by the AFOSR grant number FA9550-09-1-0583.


\end{document}